\documentclass[preprint,aip]{revtex4-1}
\usepackage{bm}
\usepackage[colorlinks=true,linkcolor=blue]{hyperref}
\usepackage[protrusion=true,expansion]{microtype}
\usepackage{amsmath}
\usepackage{amssymb}
\usepackage{amsthm}
\usepackage{amsfonts}
\usepackage{graphicx}

\usepackage{mathptmx}

\begin{document}

\title{Enhanced Spin and Electronic Reconstructions at the Cuprate-Manganite Interface}

\author{B. A.\ Gray}
\email{bagray@email.uark.edu}
\affiliation{%
Department of Physics, University of Arkansas, Fayetteville, Arkansas 70701, USA}
\author{E. J.\ Moon}
\affiliation{%
Department of Physics, University of Arkansas, Fayetteville, Arkansas 70701, USA}
\author{I. C.\ Tung}
\affiliation{%
Advanced Photon Source, Argonne National Laboratory, Argonne, Illinois 60439, USA}
\affiliation{%
Materials Science and Engineering, Northwestern University, Evanston, Illinois, 60208, USA}
\author{M.\ Kareev}
\affiliation{%
Department of Physics, University of Arkansas, Fayetteville, Arkansas 70701, USA}
\author{Jian\ Liu}
\affiliation{%
Department of Physics, University of California, Berkeley, California 94720, USA}
\author{D. J.\ Meyers}
\affiliation{%
Department of Physics, University of Arkansas, Fayetteville, Arkansas 70701, USA}
\author{M. J.\ Bedzyk}
\affiliation{%
Materials Science and Engineering, Northwestern University, Evanston, Illinois, 60208, USA}
\author{J. W.\ Freeland}
\affiliation{%
Advanced Photon Source, Argonne National Laboratory, Argonne, Illinois 60439, USA}
\author{J.\ Chakhalian}
\affiliation{%
Department of Physics, University of Arkansas, Fayetteville, Arkansas 70701, USA}

\begin{abstract}

We report on a resonant soft X-ray spectroscopy study of the electronic and magnetic structure of the cuprate-manganite interface. Polarized X-ray spectroscopy measurements taken at the Cu L edge reveal up to a five-fold increase in the dichroic signal as compared to past experimental and theoretical values. Furthermore an increase in the degree of interlayer charge transfer up to 0.25{\it e} (where {\it e} is charge of an electron) per copper ion is observed leading to a profound reconstruction in the orbital scheme for these interfacial copper ions. It is inferred that these enhancement are related to an increase in {\it T}$_{\textrm{MI}}$ observed for manganite layers grown with rapidly modulated flux.

\end{abstract}

\maketitle

\newpage 


The confluence of multiple competing order parameters stemming from the rich spectrum of ground states accessible in artificial transition metal oxide multilayers facilitates the exploration of unique quantum states and phenomena at the interface. A quintessential example is the junction between the high-$T_{\textrm{c}}$ cuprate YBa$_2$Cu$_3$O$_7$ (YBCO) and the colossal magnetoresistance manganite La$_{2/3}$Ca$_{1/3}$MnO$_3$ (LCMO), for which the exact nature of the electronic and magnetic structure at and near the interface remains actively debated. On one hand, proximity effects (PE) are excepted to play a dominate role in determining properties,\cite{Buzdin} and indeed prior extensive measurements of YBCO/LCMO heterostructures have revealed suppressions in the critical temperature and free carrier response operating over a length scale far exceeding the anticipated superconductive (SC) condensate penetration depth $\xi$$_F$.\cite{Sefrioui, Holden} One possible explanation for the observed long range PE is a triplet pairing component of the SC condensate due to the presence of magnetic inhomogeneity near the interface.\cite{Bergeret2}

Through the combined utilization of polarized neutron reflectivity and X-ray magnetic circular dichroism (XMCD), a definitive picture of the microscopic magnetic profile in YBCO/LCMO multilayers was obtained which coincided with the aforementioned theoretical predictions including an induced net moment in the SC layer at the interface with an antiferromagnetic coupling to the ferromagnetic (FM) layer.\cite{Bergeret3, Stahn, ChakhalianNat} However, subsequent X-ray linear dichroism (XLD) measurements established the role of orbital reconstruction in which a strong covalent Cu-O-Mn bond at the interface promotes charge transfer between layers and a rearrangement of the Cu $d_{z^2-r^2}$ occupation.\cite{ ChakhalianSci} The sign of the magnetic interaction between Cu and Mn is simply determined by the Goodenough-Kanamori rules, while charge transfer across the interface offers an alternative explanation to the previously seen long range PE.\cite{Holden, Freeland} Despite these advancements, recent theoretical and experimental results have reproduced the magnetic dichroic signal on Cu but failed to account for the orbital reconstruction.\cite{Andersen, Werner} Furthermore, recently evidence for a triplet SC component in YBCO/LCMO heterostructures has been emerging.\cite{Kalcheim}

In this letter, we report on a resonant X-ray absorption spectroscopy (XAS) study of the electronic and magnetic profile of the cuprate-manganite interface in the representative [YBCO (9 u.c.)/LCMO (26 u.c.)]$\times$3 superlattice (SL). Circularly polarized X-rays are used to acquire element sensitive information about the magnetic structure of the interface, while linear polarizations probe the electronic structure. Temperature dependent dc transport measurements explore the electronic and magnetic qualities of the films. We conclude that both spin and electronic reconstructions are present and markedly enhanced. Furthermore, these enhancements are linked to the structural and chemical properties of the manganite-cuprate layers obtained by rapidly modulated flux.

Sample fabrication was carried out in a recently developed Pulsed Laser Epitaxy (PLE) facility featuring infrared laser substrate heating and in-situ growth monitoring via high-pressure Reflection High Energy Electron Diffraction (HP RHEED). During deposition, stoichiometric targets of YBCO and LCMO were ablated by a KrF laser ($\lambda$ = 248 nm). The substrate temperature was held at 750 $^\circ$C, and a partial oxygen pressure of 250 mTorr was maintained inside the chamber throughout the deposition. Immediately after deposition, the substrate temperature was lowered to 530 $^\circ$C, and the sample was annealed in 500 Torr of ultra-pure oxygen for 1 hour.

A mutually compatible growth regime ({\it i.e.} temperature and pressure) for YBCO and LCMO was achieved through interval deposition in which material is deposited by high frequency pulses followed by an extended dwell time between the deposition of each unit cell (u.c.).\cite{Koster} Undamped RHEED specular intensity oscillations (not shown) were observed for both YBCO and LCMO layers over a large number of cycles allowing for a layer-by-layer growth with u.c. control. The left inset to Fig. 1 shows the post growth RHEED image for the SL along the (0 0 1) direction. The presence of unbroken crystal truncation rods up to the second order combined with well defined specular (0 0) and off specular Bragg reflections testifies to the quality of the 2-D growth.

To determine the structural properties of the SL, we performed X-ray scattering at beamline 5-BM-D of the Advanced Photon Source (APS) at Argonne National Laboratory. Figure 1 shows the specular X-ray diffraction along the (0 0 $L$) crystal truncation rod as a function of the magnitude of the out-of-plane momentum transfer vector, Q$_z$. As seen in Fig. 1, the SL shows all expected Bragg reflections for c-axis oriented layers and Kiessig fringes testifying to the quality of the film and sharpness of the interfaces. Based on the position of the YBCO (0 0 5) reflection, the average c-axis lattice constant for the YBCO layers in the SL is 11.70 \AA, which agrees with the reported bulk value for optimal stoichiometry.\cite{Ruixing}

In order to establish the electronic qualities of the samples, we investigated the dc transport properties of single layer YBCO and LCMO films and the SL. Figure 2 shows resistivity versus temperature for single layers of YBCO (left axis) and LCMO (right axis). As seen for the YBCO film, the superconducting transition $T_{\textrm{c}}$ takes place at 93 K attesting to the proper optimally doped stoichiometry. In the case of the LCMO sample, the metal-to-insulator transition takes place at {\it T}$_{\textrm{MI}}$ = 212 K (inflection point). Note, the previously reported values of {\it T}$_{\textrm{MI}}$ for LCMO films grown under similar conditions ({\it i.e.} substrate, thickness etc.) but w ithout the use of interval deposition are typically \textless 170 K.\cite{Werner} In a conventional growth to achieve an elevated $T_{\textrm{MI}}$ as in our films, significantly increased fabrication temperatures would be required which is a detrimental constraint on cuprate growth.\cite{Lebedev} The stabilization of material phases outside of regions of thermodynamic stability is a hallmark of interval deposition, and the electronic and magnetic properties of LCMO are known to depend sensitively on stoichiometry and the presence of defects.\cite{Schiffer, Aarts} Together, these help explain the origin of the enhancements seen in the spectroscopy data below. In addition, the temperature dependent resistance for the SL is plotted on the bottom left axis of Fig. 2. A suppression in the critical temperature $T_{\textrm{c}}$ = 56.7 K is observed and is most likely the result of hole depletion through interlayer charge transfer as corroborated by the XAS data below.

Next we explored the electronic and magnetic structure with XLD and XMCD resonant soft X-ray spectroscopies. The experiments were carried out at the 4-ID-C beamline of the Advanced Photon Source in Argonne National Laboratory. In the XLD studies at the Cu L edge, we investigated the orbital occupation by measuring the difference in absorption for polarizations in the ab-plane and along the c-axis, while the XMCD experiments at the Cu and Mn L edges obtained information regarding element specific magnetic moments by measuring the differences in the absorption of right and left circular polarizations. Spectra were recorded at an incident angle of 15 degrees out of plane simultaneously in both fluorescence yield (FY) and total electron yield (TEY) acquisition modes.

The linear polarization-dependent Cu L$_3$ absorption spectra of the SL is presented in Fig. 3(a). First we consider the bulk sensitive FY data set which shows the strong XLD expected for the Cu$^{2+}$ (3$d^9$) state of cuprates.\cite{Nucker, Chen} Note, the in-plane signal stems from formally divalent Cu within the CuO$_2$ sheets. Several important features can be identified in the spectrum of which the excitonic line near 931 eV is the most prominent and corresponds to transitions from the Cu 2$p$ core levels into the unoccupied bands of mainly Cu $d_{x^2-y^2}$ orbital character (2$p^6$3$d^9$ $\to$ 2$p^5$3$d^{10}$). A clear shoulder is observed on the high energy side of the white line. The shoulder is ascribed to transitions coupled with Cu ligand hole states (2$p^6$3$d^9\underline{L}$ $\to$ 2$p^5$3$d^{10}\underline{L}$) and is the well-known signature of the Zhang-Rice state.\cite{Zhang} Another important feature of the FY spectrum is connected to the c-axis polarization, where the maximum of the absorption peak is shifted by 0.3 eV to higher energy; the larger number of ligand hole states in the chains transfer spectral weight from the excitonic line to the high energy shoulder. Since the $d_{z^2-r^2}$ orbital of the CuO$_2$ planes is occupied, the out of plane polarization mainly probes unoccupied states from monovalent Cu in the chains of $d_{y^2-z^2}$ orbital character.

In the TEY data, distinct transformations in polarization dependence and lineshape manifest in the spectra indicating charge transfer across the interface leading to a reconstruction in the orbital scheme for interfacial Cu ions. As indicated in Fig. 3(a), the position of the Cu white line is shifted by 0.5 eV to lower energy in the TEY spectrum, which exceeds the value (0.4 eV) reported in the prior study. This substantial chemical shift is the mark of interfacial charge transfer. Through a comparison to Cu$^{1+}$ and Cu$^{2+}$ reference materials, an approximate calculation yields a charge transfer amplitude of 0.25{\it e} per copper ion. \cite{deGroot} Furthermore, the shoulder at high energy is not observed in the TEY data confirming the depletion of holes from the YBCO region of the interface. While the evolution in the amplitude of the high energy shoulder is a common feature of YBCO XAS doping profiles, the shift in the white line is unexpected. \cite{Nucker} Previously, this has been attributed to the reconstruction of the $d_{z^2-r^2}$ orbital through the formation of a covalent bond with Mn through apical oxygen across the interface.\cite{ChakhalianSci, Liu} Confirming this picture, the large linear dichroism of the bulk is no longer present, and the absorption along the c-axis now even surpasses that of the the in-plane for the TEY spectrum.

Finally, we turn our attention to magnetism on Cu and Mn. Figure 3(b) shows the XMCD spectra at both the Mn and Cu L-edges acquired in TEY mode. As anticipated for a FM system, the strong magnetic dichroism reaches a maximum value of 37.4 $\%$ at the Mn L$_3$ edge. On the Cu L-edge, however, only a moderate magnetic signal is present, which indicates an uncompensated magnetic moment on Cu near the interface. From the sign of the XMCD on both Mn and Cu edges, an antiparallel alignment of Cu and Mn moments coupled across the interface can be deduced. The magnitude of the dichroism (6.9 $\%$) a the Cu L$_3$-edge is up to a {\it factor five larger} than the previously reported values providing additional evidence for the high quality of the interface.\cite{ChakhalianNat} The measurements at the Cu L-edge were repeated at 100 mT without loss in the magnitude of the magnetic dichroism.

In summary, our resonant soft X-ray spectroscopy study of the cuprate-manganite interface has confirmed the presence of both spin and orbital reconstructions. We observed increases in the dichroic signal and the chemical shift at the Cu L edge signaling an enhancement in the induced moment on Cu and the degree of interlayer charge transfer, respectively. Furthermore, an increase in {\it T}$_{\textrm{MI}}$ has been observed for manganite layers grown with interval deposition, which is fundamentally related to its magnetic and electronic properties.

J.C. was supported by the DOD-ARO under Grant No. \ 0402-17291 and the NSF under Grant No.\ DMR-0747808. Work at the APS is supported by the U.S.\ DOE under grant no.\ DEAC02-06CH11357.

\newpage

\begin{figure}[t]\vspace{-0pt}
 \includegraphics[width=11cm]{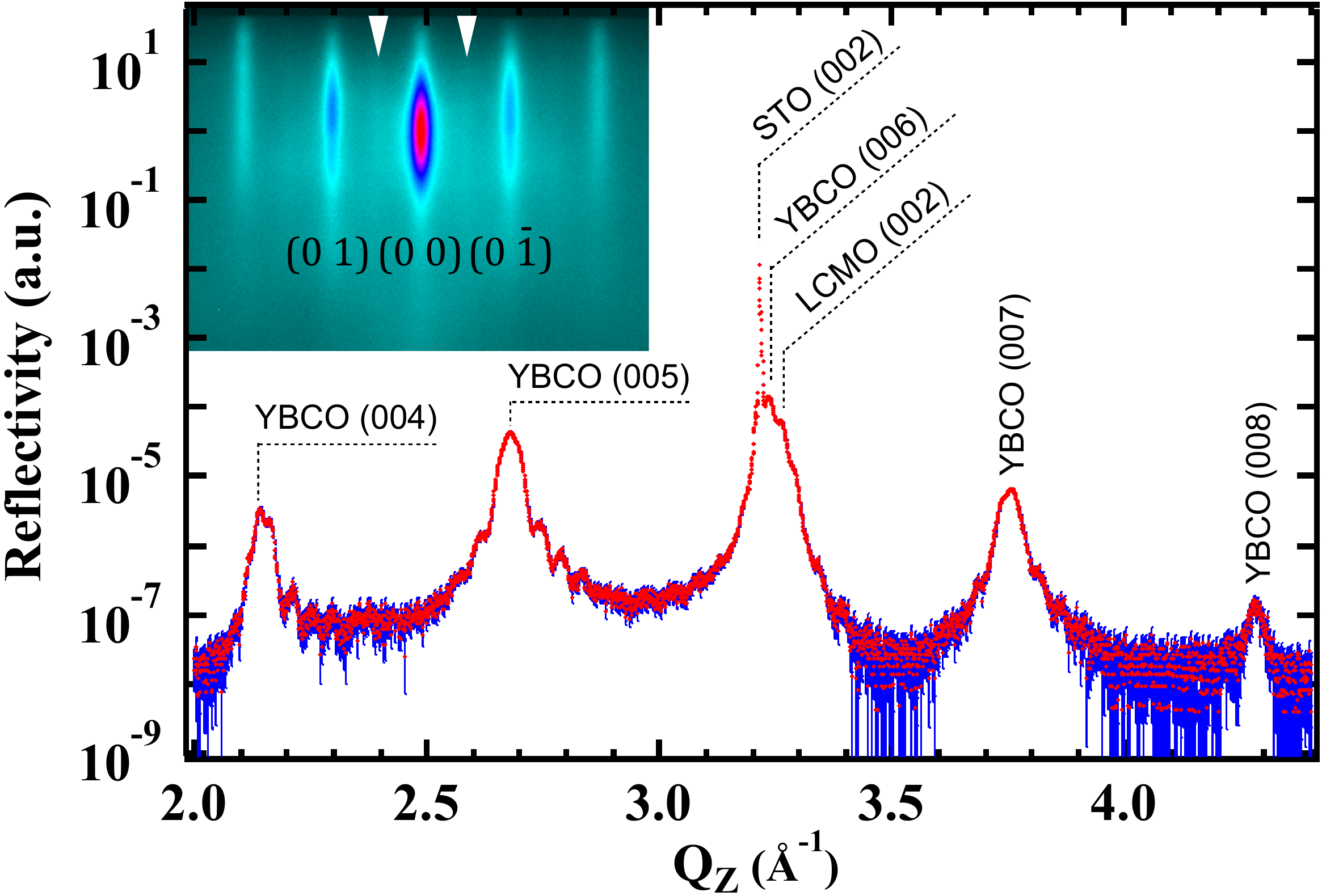}
   \caption{\label{Diffraction}(Color online)
   X-ray reflectivity data (dots) and error bar (solid line) for a [YBCO (9 u.c.)/LCMO (26 u.c.)]$\times$10 SL SrTiO$_3$. The inset on the left shows the after growth RHEED image for the same SL on a SrTiO$_3$ substrate. White triangular markers indicate the weak half-order signal which is attributed to the {\it Pbnm} crystal symmetry of the capping LCMO layer. The right-hand inset is a high angle annular dark field scanning transmission electron microscopy image of the YBCO/LCMO interfacial region. Labels indicate different layers.}
\end{figure}

\begin{figure}[t]\vspace{-0pt}
 \includegraphics[]{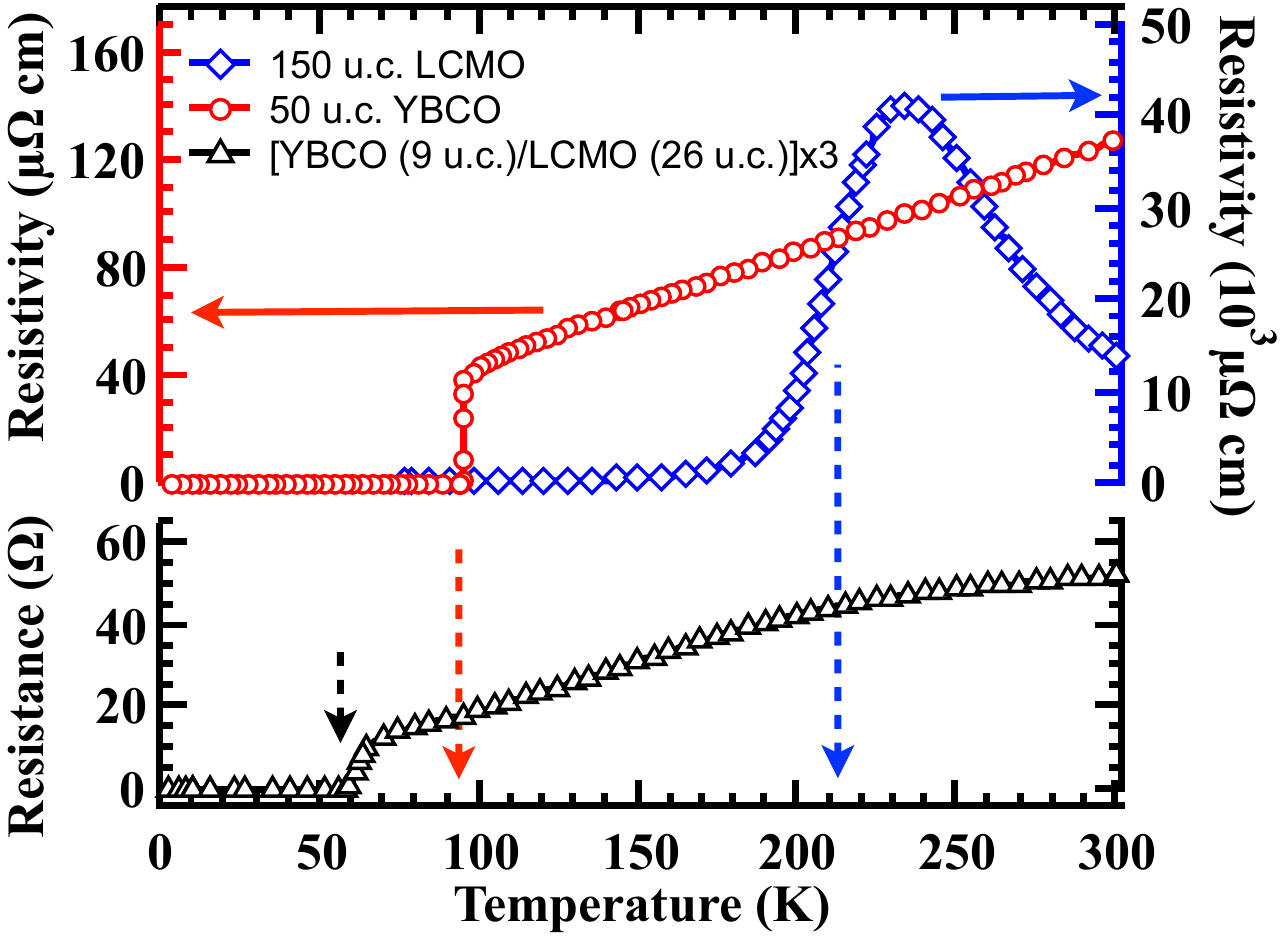}
   \caption{\label{Transport}(Color online)
    Temperature-dependent dc transport for 50 u.c. YBCO and 150 u.c. LCMO single layers and the [YBCO (9 u.c.)/LCMO (26 u.c.)]$\times$3 SL on SrTiO$_3$ substrates. Measurements were performed in the conventional {\it van der Paw} configuration.}
\end{figure}

\begin{figure}[t]\vspace{-0pt}
 \includegraphics[]{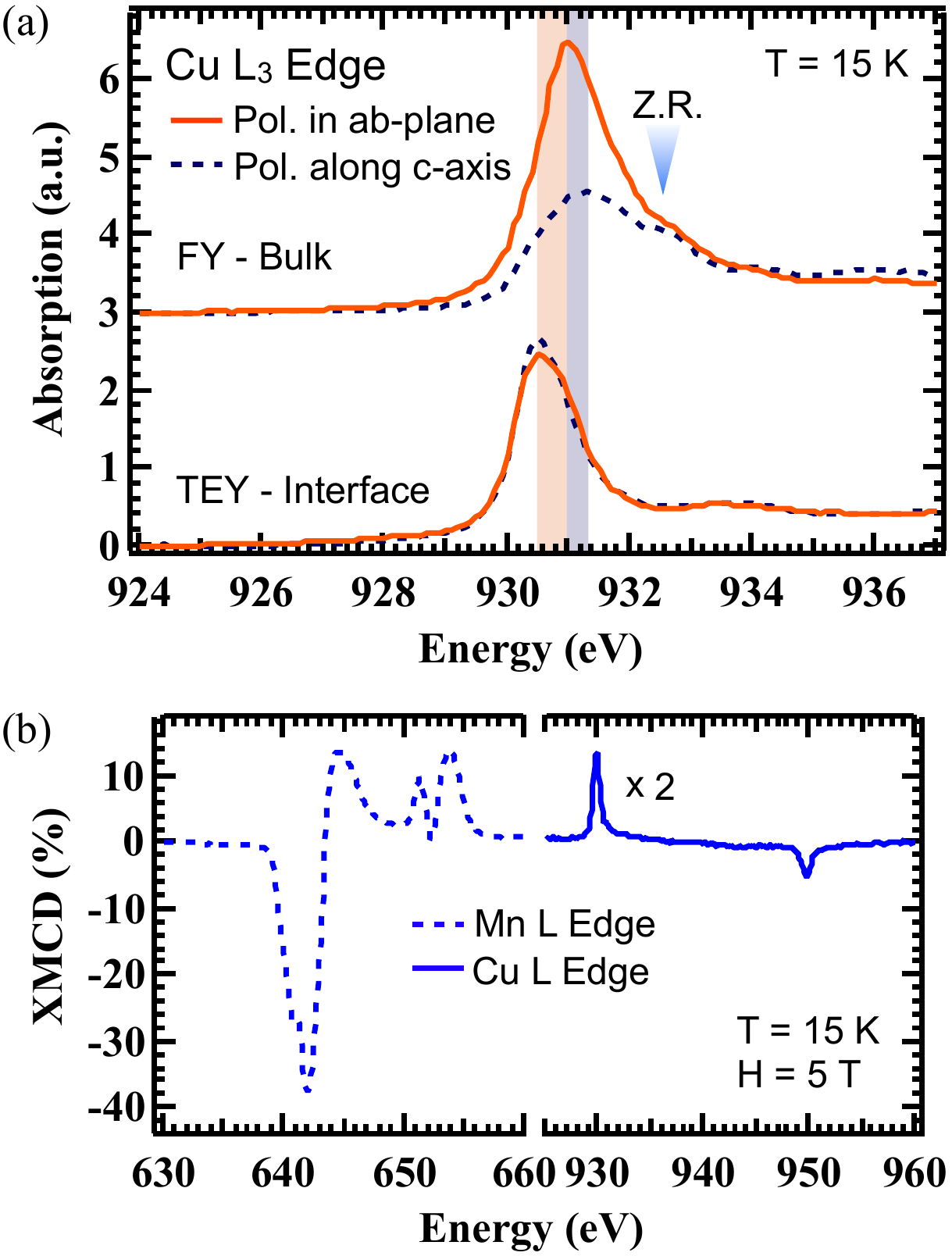}
   \caption{\label{Spec}(Color online)
   Soft X-ray spectroscopies of the [9 u.c. YBCO/ 26 u.c. LCMO]$\times$3 SL on SrTiO$_3$. (a) XLD measurements at the Cu L$_3$-edge. The top and bottom sets of spectra correspond to FY and TEY detection modes, respectively. Due to TEY's shallow probing depth, only Cu at the first interface predominantly contributes to the Cu L intensity in this mode. (b) XMCD measurements at the Mn (left) and Cu (right) L-edges in TEY. Note, to rule out artifacts the XMCD data sets were checked at both magnetic field orientations.}
\end{figure}

\end{document}